\documentclass[twocolumn]{aastex631}

\shorttitle{The False Spin of an Exo-Venus}
\shortauthors{Stephen R. Kane}

\begin{document}

\title{The False Spin of an Exo-Venus}

\author[0000-0002-7084-0529]{Stephen R. Kane}
\affiliation{Department of Earth and Planetary Sciences, University of
California, Riverside, CA 92521, USA}
\email{skane@ucr.edu}


\begin{abstract}

Direct imaging of terrestrial exoplanets will enable rotational and
atmospheric characterization through time-resolved photometry and
high-dispersion spectroscopy. However, the velocity field inferred
from reflected light does not necessarily correspond to the rotation
of the solid planet, but rather to the motion of the layer from which
the photons emerge. Venus provides a crucial Solar System example of
this ambiguity: the solid planet rotates slowly, whereas the
cloud-level atmosphere exhibits superrotation with a period of only
several days. Here we investigate the observational degeneracy between
rapid planetary rotation and atmospheric superrotation. We construct a
disk-integrated reflected-light velocity model that includes
solid-body rotation, zonal winds, and phase-dependent illumination. We
show that, for a single spectral tracer probing a narrow range of
pressures, a zonal wind field whose latitude dependence is similar to
solid-body rotation can exactly mimic the line profile of a rapidly
rotating planet. The degeneracy can be broken by measuring the
apparent rotational velocity as a function of wavelength or line
formation pressure. For a Venus-like wind profile, the apparent period
can vary from hundreds of days in the lower atmosphere to
$\sim$4--5~days at the cloud deck. We estimate the resolving power and
signal-to-noise ratio required to measure this vertical shear. The
most robust diagnostic of atmospheric superrotation is not a single
value of $v \sin i$, but an altitude-dependent ``false spin''
signature across multiple spectral tracers. These results have direct
implications for interpreting rotational measurements of Venus-like
worlds with the Habitable Worlds Observatory and complementary
high-dispersion facilities.

\end{abstract}



\section{Introduction}
\label{sec:intro}

The direct imaging of terrestrial exoplanets is a major long-term goal
of exoplanet science, motivated by the need to detect and characterize
temperate rocky planets around nearby stars. Mission studies such as
HabEx and LUVOIR, and ongoing preparatory work for the Habitable
Worlds Observatory (HWO), have emphasized the need for precursor
science that will inform target selection, observing strategies, and
atmospheric interpretation
\citep{reportluvoir,reporthabex,harada2024b,kane2024d,kane2024e,stark2024b,harada2025}. These
studies generally focus on the detection criteria for planets that
reside within the Habitable Zone (HZ) of their stars
\citep{kasting1993a,kane2012a,kopparapu2013a,kopparapu2014,kane2016c,hill2018,hill2023}. However,
within that context, planets similar to Venus are not merely
contaminants to a search for Earth analogs. Rather, they are essential
test cases for interpreting terrestrial planet spectra, clouds,
atmospheric evolution, and the inner edge of the HZ
\citep{kane2014a,kane2021a,kane2026a}.

A common objective of time-resolved direct imaging is to determine the
planetary rotation period. Photometric modulation by surface features
or clouds, phase-dependent albedo maps, and rotational broadening in
high-resolution spectra each offer a route to the spin state. The
method is powerful because the rotation period of a planet encodes its
formation history, tidal evolution, atmospheric dynamics, and
potential climate regime
\citep{kawahara2010,fujii2012,cowan2013a}. Ground-based and
space-based studies have demonstrated that time-resolved reflected
light can be inverted to recover latitudinal and longitudinal surface
maps via spin-orbit tomography
\citep{cowan2009,kawahara2010,kawahara2011,fujii2012,fujii2017c}, and
rotation period detection has been explored for Earth-like exoplanets
at a range of signal-to-noise ratios \citep{li2022a}. Future
instruments will extend these techniques to HZ rocky planets
\citep{snellen2015,lovis2017,wang2017a}. The danger is that the
quantity being measured is not always the spin of the solid body. In
reflected light, the observable signal is determined by the altitude
of the scattering or absorbing layer. A cloudy terrestrial planet may
therefore reveal the dynamics of its cloud tops rather than its
surface.

Venus provides the warning case. The sidereal rotation period of the
solid planet is 243 days, yet the cloud-level atmosphere superrotates
with a characteristic period of several days. Observations from
Akatsuki have shown that disk-integrated brightness modulations of
Venus occur at 3.7 and 4.6 days, reflecting atmospheric waves and
superrotation rather than the solid-body rotation period
\citep{lee2020b}. Thus, Venus can appear to be a false positive for
rotation measurements based on unresolved photometric variability,
which is also the ambiguity identified for Earth analogs by
\citet{palle2008} and quantified via spin-orbit tomography by
\citet{kawahara2011}. This result naturally raises a spectroscopic
version of the same problem: if a directly imaged exoplanet exhibits
broad or shifted spectral lines, how does one distinguish a rapidly
rotating planet from a slowly rotating planet with a superrotating
atmosphere?

High-dispersion spectroscopy already provides a means of measuring
velocity fields in exoplanet atmospheres. The rotational broadening of
the directly imaged giant planet $\beta$~Pictoris~b was measured from
high-resolution molecular spectra \citep{snellen2014b}, and related
work has explored Doppler imaging of substellar atmospheres
\citep{crossfield2014b}. For transiting hot Jupiters, both rotation
and winds can broaden, shift, and distort high-resolution spectral
lines
\citep{showman2013a,kempton2014b,brogi2016,seidel2020a}. Reflected-light
high-resolution spectroscopy further demonstrates that line broadening
can substantially affect the detectability and interpretation of
planetary signals \citep{spring2022,winterhalder2026b}. Recent
high-resolution observations of directly imaged planets likewise show
that molecular-line broadening can constrain planetary spin
\citep{snellen2014b,parker2024}. The rotation rate of a transiting
planet can also be constrained from the shape of its transit spectrum
\citep{spiegel2007}. The combination of high-dispersion spectroscopy
with high-contrast imaging provides a particularly promising route to
applying these methods to directly imaged rocky planets
\citep{snellen2015,wang2017a}, with ongoing developments targeting
ELT-class facilities \citep{lovis2017}.

In this paper, we identify the rotation-wind degeneracy in its
simplest form and determine which observations can break that
degeneracy. We focus on Venus-like terrestrial planets as a case study
because their thick cloud decks make the observable layer especially
decoupled from the surface. Section~\ref{sec:model} describes the
disk-integrated velocity model. Section~\ref{sec:results} presents the
degeneracy and its wavelength-dependent diagnostic. We discuss the
implications for direct imaging of Venus-like worlds in
Section~\ref{sec:disc}, and provide concluding remarks in
Section~\ref{sec:con}.


\section{Reflected-Light Velocity Model}
\label{sec:model}

The relevant observable for high-dispersion reflected-light
spectroscopy is the disk-integrated line profile of the planet,
following the reflected-light disk-integration geometry described by
\citet{crossfield2014b} and \citet{spring2022}. We
model the planet as a sphere of radius $R_p$, observed at orbital phase
angle $\alpha$, where $\alpha=0^\circ$ corresponds to full phase. The
spin axis is assumed to be perpendicular to the orbital plane, and the
planet is viewed edge-on. These assumptions maximize the projected
velocity and are adopted to isolate the rotation-wind degeneracy,
rather than to describe the full range of possible obliquities and
viewing geometries. For a general spin-axis inclination $i_{\rm spin}$,
the measured velocity width scales with $\sin i_{\rm spin}$,
so absolute period determinations require knowledge of the viewing
geometry. The vertical-shear diagnostic, however, is preserved because
all atmospheric layers share the same projection factor, so the ratio
of apparent velocities between layers is independent of
$i_{\rm spin}$.

For a surface element at latitude $\phi$ and longitude $\lambda$,
where $\lambda=0$ is the sub-observer longitude (the center of the
visible disk), the local
line-of-sight velocity is
\begin{equation}
v_{\rm los}(\phi,\lambda,p) =
-\left[\frac{2\pi R_p}{P_{\rm rot}}\cos\phi +
u_\phi(\phi,p)\right]\sin\lambda ,
\label{eq:vlos}
\end{equation}
where $P_{\rm rot}$ is the solid-body rotation period and
$u_\phi(\phi,p)$ is the zonal wind velocity at the pressure level
$p$ probed by the observation. Equation~(\ref{eq:vlos}) is written in
terms of the velocity of the layer that produces the reflected or
absorbed photons. The sign convention is arbitrary for the present
analysis, since the primary diagnostic considered here is the
magnitude and shape of the velocity kernel (the distribution of
line-of-sight velocities weighted by the reflected-light contribution
across the visible disk, defined formally in
Equation~\ref{eq:kernel}). Venus rotates retrograde,
and the cloud-level superrotation moves in the same direction as the
solid body but roughly 60 times faster; because
the solid-body equatorial velocity is only 1.81~m~s$^{-1}$ compared
with cloud-top winds near 100~m~s$^{-1}$, the numerical examples
below use the magnitude of $u_0$ and the distinction between prograde
and retrograde is negligible for the present analysis.

The reflected-light contribution from each surface element is weighted
by a Lambertian illumination and visibility function,
\begin{equation}
  W(\phi,\lambda,\alpha) =
  \mu_{\rm obs}\,\mu_\star\,
  H(\mu_{\rm obs})\,H(\mu_\star),
  \label{eq:weight}
\end{equation}
where $H$ is the Heaviside function,
$\mu_{\rm obs}=\cos\phi\cos\lambda$, and
$\mu_\star=\cos\phi\cos(\lambda-\alpha)$. The disk-integrated velocity
kernel for a layer at pressure $p$ is then
\begin{equation}
  K(v,p,\alpha) =
  \int W(\phi,\lambda,\alpha)
  \delta\left[v-v_{\rm los}(\phi,\lambda,p)\right]
  \cos\phi\,d\phi\,d\lambda ,
  \label{eq:kernel}
\end{equation}
where the additional factor of $\cos\phi$ is the surface area element.

If the zonal wind has the form
\begin{equation}
  u_\phi(\phi,p) = u_0(p)\cos\phi ,
  \label{eq:wind}
\end{equation}
then Equation~(\ref{eq:vlos}) becomes
\begin{equation}
  v_{\rm los} =
  -\left[\frac{2\pi R_p}{P_{\rm rot}}+u_0(p)\right]
  \cos\phi\,\sin\lambda .
  \label{eq:degenerate}
\end{equation}
Thus, at a single pressure level and a single wavelength, a
superrotating atmosphere with this latitude dependence is exactly
degenerate with a solid body rotating at an effective equatorial
velocity
\begin{equation}
  v_{\rm app}(p) =
  \frac{2\pi R_p}{P_{\rm rot}}+u_0(p).
  \label{eq:vapp}
\end{equation}
The corresponding apparent rotation period is
\begin{equation}
  P_{\rm app}(p) = \frac{2\pi R_p}{|v_{\rm app}(p)|}.
  \label{eq:papp}
\end{equation}

For the numerical examples below, we adopt the radius of Venus
($R_p=6051.8$~km; for a cloud-top scattering layer the effective
radius is $R_p + z_{\rm cloud}$, but the correction is $\sim 1$\% for
Venus and is neglected here) and a Venus solid-body rotation period of
243.025~days, corresponding to an equatorial velocity of
1.81~m~s$^{-1}$. A cloud-top velocity of 100~m~s$^{-1}$ corresponds to
an apparent period of 4.4~days. The vertical wind profile used in this
work is a simple constructed profile, not digitized from a specific
measurement, with $u_0\approx 1$~m~s$^{-1}$ near the surface and
$u_0\approx 100$~m~s$^{-1}$ near the cloud deck. It is designed to
approximate the order-of-magnitude vertical structure of Venus as
measured by entry probes and cloud tracking, in which the zonal wind
increases from near-surface values below 5~m~s$^{-1}$ to cloud-top
velocities near 100~m~s$^{-1}$ at $\sim 70$~km altitude
\citep{khatuntsev2013,bertaux2016,horinouchi2020}. The atmospheric
dynamics and superrotation of Venus have been extensively studied
observationally and theoretically
\citep{lee2007c,lebonnois2010,showman2011b,peralta2017a,limaye2018a,read2018b,horinouchi2020,imamura2020,lebonnois2020b,cohen2024c},
and the profile adopted here captures the essential dynamical
structure for the present illustrative purpose rather than serving as
a full Venus general circulation model.


\section{Results}
\label{sec:results}


\subsection{A Single-Layer Degeneracy}
\label{sec:singlelayer}

\begin{figure}
  \centering
  \includegraphics[angle=270,width=\columnwidth]{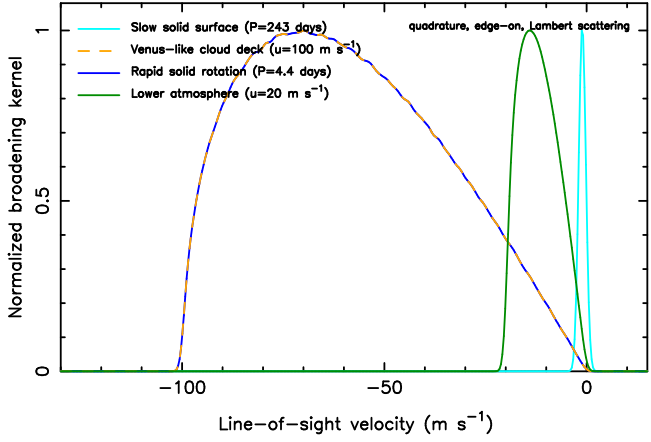}
  \caption{Disk-integrated velocity kernels at quadrature for several
    idealized terrestrial planets. The slow solid case adopts the
    solid-body rotation of Venus ($P=243$~days). The Venus-like cloud
    deck assumes the same slow solid-body rotation but adds a
    cloud-level zonal velocity of 100~m~s$^{-1}$. The rapid solid-body
    model has the same apparent equatorial velocity but no atmospheric
    wind. The Venus-like cloud deck and rapid solid rotation cases
    produce mathematically identical kernels (the orange dashed curve
    overlays the blue solid curve exactly), demonstrating the exact
    degeneracy proved in Equation~(\ref{eq:degenerate}) for a
    $\cos\phi$ zonal wind profile. The lower atmosphere case
    ($u=20$~m~s$^{-1}$) produces a distinct, narrower kernel that is
    clearly separated from the degenerate pair.}
  \label{fig:profiles}
\end{figure}

Figure~\ref{fig:profiles} shows the disk-integrated velocity kernels
at quadrature for four illustrative cases. At quadrature, only a
portion of the illuminated disk is visible to the observer, so the
velocity kernels include both a broadening component and a
phase-dependent centroid shift; they are not symmetric about zero
velocity as would be the case for a uniformly illuminated disk viewed
at full phase. A slowly rotating solid
planet with no atmospheric motion produces a narrow kernel. A
lower-atmosphere layer with a modest zonal wind of
20~m~s$^{-1}$ produces a broader and shifted kernel due to the
phase-dependent illumination of the rotating or moving layer. The
crucial comparison is between the Venus-like cloud deck and a rapidly
rotating solid planet. These two profiles are indistinguishable in the
case where the cloud-level zonal wind has the latitude dependence of
Equation~(\ref{eq:wind}). In other words, the reflected-light spectrum
of a slowly rotating planet with a superrotating cloud deck can return
the same apparent velocity kernel as a solid planet rotating with a
period of several days.

This result holds for any spectral tracer that probes the same
atmospheric layer. A line formed in the cloud deck, a reflected
stellar line scattered by the cloud deck, or a molecular line whose
contribution function peaks in the same atmospheric layer will all
inherit the same velocity field. Therefore, the interpretation of
$v\sin i$ for a cloudy terrestrial exoplanet must be explicitly tied
to the pressure level or scattering altitude of the observation, a
challenge that compounds the already difficult problem of mapping
cloudy surfaces from unresolved photometry \citep{teinturier2022}.

The degeneracy in Equation~(\ref{eq:degenerate}) is exact only for the
idealized $\cos\phi$ latitude dependence of
Equation~(\ref{eq:wind}). Real Venus GCMs and observations reveal that
the actual zonal wind profile departs significantly from $\cos\phi$:
the cloud-level circulation exhibits a broad equatorial jet, distinct
mid-latitude jets near $\pm 45^{\circ}$, and strong polar vortices
\citep{khatuntsev2013,bertaux2016,horinouchi2020}. These departures
from solid-body latitude structure would produce asymmetries in the
disk-integrated line profile that could, in principle, break the
degeneracy even at a single pressure level given sufficient
signal-to-noise ratio. The $\cos\phi$ case therefore represents a
worst-case scenario for distinguishing atmospheric superrotation from
rapid solid-body rotation, and the true diagnostic potential of
high-resolution spectroscopy is correspondingly greater than the
present model suggests.


\subsection{Vertical Shear as the Primary Diagnostic}
\label{sec:vertical}

\begin{figure}
  \centering
  \includegraphics[angle=270,width=\columnwidth]{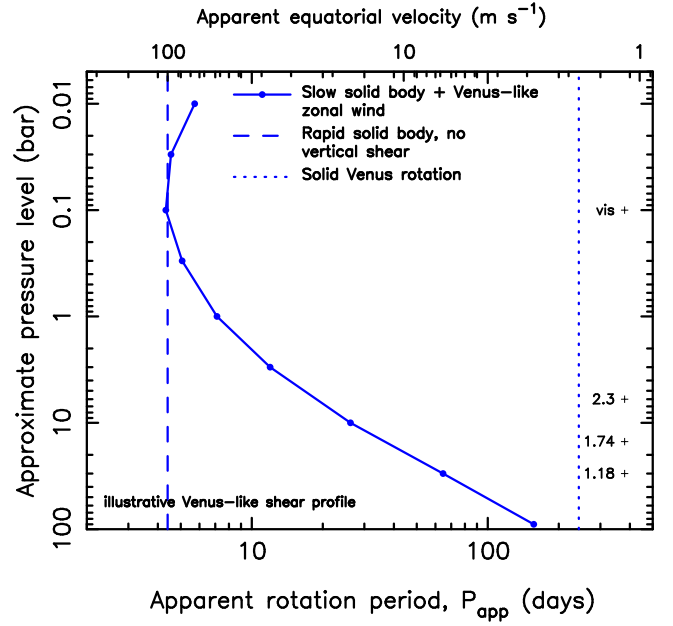}
  \caption{Apparent rotation period as a function of pressure level
    for an illustrative Venus-like vertical wind profile. The blue
    curve shows a slowly rotating solid planet with altitude-dependent
    zonal winds. The dashed line shows a rapidly rotating solid planet
    with the same cloud-level apparent velocity but no vertical
    shear. The dotted line shows the true solid-body rotation period
    of Venus. The upper axis shows the corresponding apparent
    equatorial velocity, which scales inversely with the period.
    Point markers in the right margin indicate the approximate
    pressure levels probed at visible wavelengths (``vis'', cloud
    deck, $\sim 0.1$~bar) and in the near-infrared thermal windows at
    2.3, 1.74, and 1.18~$\mu$m ($\sim 6$, 15, and 30~bar,
    corresponding to $\sim 35$, 25, and 15~km altitude respectively),
    illustrating the range of atmospheric depths accessible to
    multi-wavelength velocity measurements. The slight reversal in
    $P_{\rm app}$ at low pressures (above $\sim 0.03$~bar) reflects
    the decrease in zonal wind speed above the cloud deck in the
    adopted profile, and is not a plotting artifact. A wavelength- or
    line-strength-dependent apparent rotational velocity is the
    primary diagnostic of atmospheric superrotation.}
  \label{fig:shear}
\end{figure}

The strongest discriminator between rapid solid-body rotation and
atmospheric superrotation is the variation of $P_{\rm app}$ with
pressure level. Figure~\ref{fig:shear} shows the apparent rotation
period inferred from the illustrative Venus-like wind profile. A
rapidly rotating solid planet without vertical shear produces the same
apparent period at all altitudes. A slowly rotating Venus-like planet
instead produces a continuum of apparent periods, from
$\sim 150$~days in the lowest atmospheric layer represented here to
$\sim 4$--5~days near the cloud deck.

The practical implication is that no single reflected-light detection
should be interpreted as the spin period of the underlying planet
without additional information. Instead, the observable of interest is
the vector of apparent velocities measured from different spectral
regions:
\begin{equation}
  \mathbf{v}_{\rm app} = \left(v_{\rm app,1},v_{\rm app,2},...,v_{\rm
    app,N}\right),
  \label{eq:vvector}
\end{equation}
where each component corresponds to a different wavelength interval,
molecule, line strength, or contribution function. A rapidly rotating
solid planet predicts $\mathbf{v}_{\rm app}$ to be approximately
constant, after accounting for known changes in planetary radius and
viewing geometry. A superrotating atmosphere predicts an altitude
sequence.

For Venus analogs, the most promising comparison is between reflected
stellar lines dominated by the upper cloud deck and spectral windows
or molecular line wings that probe deeper layers. In the visible,
cloud opacity may prevent direct access to the lower atmosphere. In
the near-infrared, thermal emission windows at 1.18, 1.74, and
2.3~$\mu$m provide direct access to the sub-cloud atmosphere of Venus
through gaps in the CO$_2$ opacity
\citep{meadows1996,arney2014,taylor2018}. However, thick cloud decks
can mute spectral features and hamper atmospheric detections even at
high dispersion. For Venus itself, the near-infrared night-side
windows correspond to the wavelengths at which this muting is
minimized: the 2.3, 1.74, and 1.18~$\mu$m windows probe progressively
deeper atmospheric layers, with contribution functions peaking near
35, 25, and 15~km altitude ($\sim 6$, 15, and 30~bar) respectively
\citep{meadows1996,marcq2018,taylor2018}. Thermal emission emerging
through these windows has been used to map the deep atmospheric
circulation of Venus from ground-based and orbital observations,
demonstrating that sub-cloud velocity information is observationally
accessible in at least this Solar System case despite the overlying
cloud deck. The degree to which analogous windows exist and remain
uncontaminated in an exo-Venus atmosphere will depend on its specific
cloud and haze structure, which may differ substantially from
Venus. These windows are strictly thermal-emission features, most
readily observed on the nightside, and would require an emitted-light
version of Equation~(\ref{eq:kernel}) with a different disk-weighting
function than the reflected-light Lambertian model adopted here. They
are invoked as the natural lower-atmosphere analog because they
represent the only demonstrated means of spectroscopically accessing
sub-cloud velocities on Venus, and analogous windows in exo-Venus
spectra could provide the pressure leverage needed to detect vertical
shear.  Additionally, because the opacity at line center is much
larger than in the wings, the atmosphere becomes optically thick to
line-center photons at a higher altitude (lower pressure) than to
photons in the line wings. Comparing the apparent velocity derived
from a strong-line core to that from a weak line or line wing
therefore probes a range of contribution function altitudes within a
single spectral bandpass. The diagnostic is thus conceptually clean
but instrumentally demanding.


\subsection{Phase-Resolved Profiles}
\label{sec:phase}

\begin{figure}
  \centering
  \includegraphics[angle=270,width=\columnwidth]{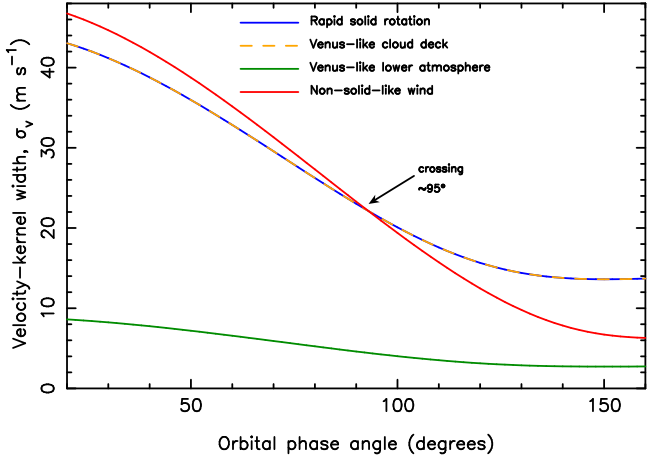}
  \caption{Velocity-kernel width as a function of orbital phase. The
    rapidly rotating solid planet and the Venus-like cloud deck
    overlap because their line-of-sight velocity fields are
    mathematically equivalent at a single pressure level
    (Equation~\ref{eq:degenerate}). A lower atmosphere with weaker
    winds ($u=20$~m~s$^{-1}$) is clearly separated at all phases. The
    non-solid-like wind profile ($u_\phi \propto \cos^{0.35}\phi$,
    representing a wind that extends more broadly in latitude
    than solid-body rotation) crosses the degenerate pair near
    $\sim 95^{\circ}$ phase, demonstrating that sufficiently precise
    phase-resolved spectra can constrain the latitudinal structure of
    atmospheric flow.}
  \label{fig:phase_widths}
\end{figure}

Phase-resolved spectroscopy provides additional information because
the visible and illuminated portions of the velocity field change with
orbital phase. Figure~\ref{fig:phase_widths} shows the phase
dependence of the velocity-kernel width $\sigma_v$, defined as the
flux-weighted standard deviation of the kernel $K(v)$, for four
cases. The rapidly rotating solid planet and the Venus-like cloud deck
remain degenerate when their velocity fields share the form of
Equation~(\ref{eq:degenerate}). A lower-atmosphere layer has a smaller
kernel width at all phases. A wind profile that does not follow the
solid-body latitude dependence produces a notably different phase
dependence; in the ``non-solid-like wind'' case ($u_\phi \propto
\cos^{0.35}\phi$, representing a wind that extends more broadly in
latitude than $\cos\phi$, with relatively stronger contributions at
mid-latitudes and weaker equatorial confinement), the kernel width
crosses the degenerate pair near $\sim 95^{\circ}$ phase. The exponent
$q=0.35$ is adopted as an illustrative departure from solid-body
rotation; its specific value is not derived from a particular
atmospheric model, but the broader-than-$\cos\phi$ latitude structure
it produces is qualitatively consistent with Venus cloud-tracked wind
measurements, which show relatively uniform zonal velocities from the
equator to mid-latitudes before decreasing poleward
\citep{khatuntsev2013,horinouchi2020}. The crossing near $\sim
95^{\circ}$ occurs because the broader latitude distribution of the
wind is more efficiently projected at lower phase angles but diluted
at larger phase angles as the visible illuminated crescent narrows and
shifts toward the sub-stellar longitude. Such signatures suggest that
sufficiently precise phase-resolved spectra could constrain the
latitudinal structure of the atmospheric flow and partially break the
degeneracy even at a single pressure level.

Realizing this phase-resolved diagnostic in practice faces two
significant observational challenges. First, distinguishing the
non-solid-like wind case from the degenerate pair requires sampling
the kernel width across a phase range that brackets the $\sim
95^{\circ}$ crossing; a minimum of three to four epochs spanning
roughly $40^{\circ}$--$140^{\circ}$ in phase angle would be needed to
establish the differing phase trends with any confidence, and finer
sampling would be required to characterize the latitudinal structure
in detail. Because exo-Venus targets will be extremely faint,
achieving the necessary S/N at each epoch may require stacking
multiple deep integrations, which limits the effective time resolution
and favors targets whose apparent rotation period is long enough that
the atmospheric state does not evolve substantially between stacked
exposures. Second, assigning the correct phase to each spectrum
requires a known ephemeris. For the orbital phase angle $\alpha$, this
is set by the planet's orbit and is generally well determined. For the
rotational or wind phase, however, a precise photometric period would
in most cases be a prerequisite, since the several-day apparent
periods of these atmospheres make it difficult to assign a consistent
rotational phase across epochs separated by many rotation cycles
without an independent period measurement. This requirement reinforces
the value of the combined photometric-plus-spectroscopic strategy
discussed in Section~\ref{sec:combined}. We emphasize that these
challenges apply specifically to the phase-resolved morphology
diagnostic; the primary vertical-shear diagnostic
(Section~\ref{sec:vertical}) does not require phase assignment or a
temporal baseline, since it relies on comparing velocity widths across
wavelengths at a single epoch.

\begin{figure}
  \centering
  \includegraphics[angle=270,width=\columnwidth]{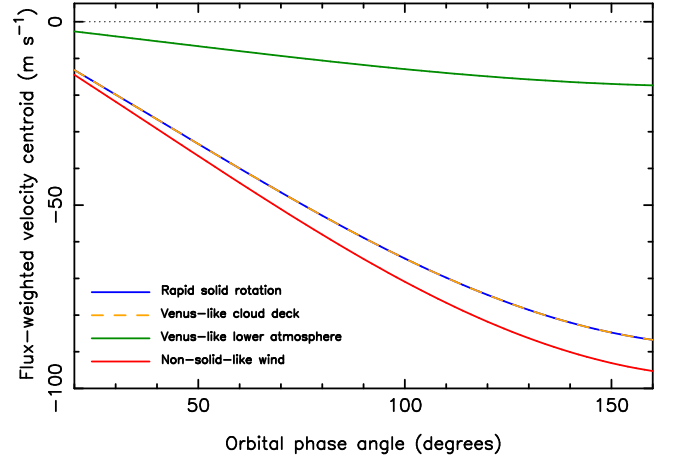}
  \caption{Flux-weighted velocity centroid as a function of orbital
    phase for the same cases as Figure~\ref{fig:phase_widths}. The
    phase range begins at $20^{\circ}$ to avoid the near-full-phase
    regime where the centroid approaches zero by symmetry. The
    centroid shifts arise from the phase-dependent illumination and
    visibility of the moving layer. The Venus-like cloud deck and
    rapid solid rotation cases produce identical centroid curves (the
    orange dashed curve lies directly beneath the blue solid curve).
    Phase-dependent centroid offsets of order tens of m~s$^{-1}$ can
    contaminate orbital radial velocity measurements and must be
    modeled when interpreting reflected-light spectroscopy of
    Venus-like planets. The non-solid-like wind (broader latitude
    wind distribution) produces a systematically more negative
    centroid at large phase angles, a consequence of the enhanced
    mid-latitude contribution from the wind.}
  \label{fig:phase_centroids}
\end{figure}

Figure~\ref{fig:phase_centroids} shows the corresponding flux-weighted
velocity centroids. The centroids are not generally zero because
reflected light preferentially weights different limbs of the planet
at different phases. This phase-dependent centroid should not be
mistaken for an orbital radial velocity systematic error. It is
instead a rotational or atmospheric analog of a
Rossiter--McLaughlin-like weighting of the visible planetary disk
\citep{queloz2000b,gaudi2007}.  The phase-dependent centroid shift
amounts to several tens of m~s$^{-1}$ at large phase angles for the
100~m~s$^{-1}$ cases, which is at the precision frontier of the most
sensitive current and planned spectrographs \citep[e.g., ESPRESSO,
  ELT/ANDES, VLT/RISTRETTO;][]{pepe2021,lovis2022,marconi2024a} and
could therefore contaminate orbital radial velocity measurements if
not properly modeled. As with the kernel width, a solid-like
superrotating wind remains degenerate with rapid solid-body rotation,
whereas vertical shear or non-solid-like latitude structure can break
the degeneracy.

\subsection{Detectability of Vertical Shear}
\label{sec:detectability}

\begin{figure}
  \centering
  \includegraphics[angle=270,width=\columnwidth]{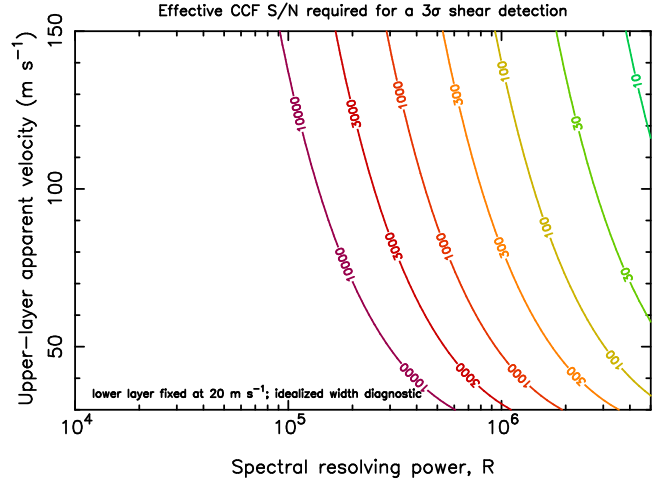}
  \caption{Idealized effective CCF signal-to-noise ratio required for
    a 3$\sigma$ detection of vertical shear, using
    Equation~(\ref{eq:snr}). The lower layer is fixed at an apparent
    velocity of 20~m~s$^{-1}$, while the upper-layer velocity and
    spectral resolving power are varied. Contour colors indicate the
    required S/N, progressing from green (favorable, S/N $\lesssim 30$)
    through gold and orange to red and purple (demanding, S/N
    $\gtrsim 10^3$). The blank lower-left region corresponds to
    parameter space where the required S/N exceeds the plotted range.
    The figure illustrates the severe resolution dependence of
    terrestrial wind measurements.}
  \label{fig:detectability}
\end{figure}

A complete detectability calculation requires a full model of the
planet spectrum, stellar leakage, speckle chromaticity, detector
noise, telluric contamination for ground-based observations, and the
wavelength-dependent planet/star contrast. Following the
cross-correlation framework developed for high-dispersion coronagraphy
\citep{snellen2015,wang2017a,spring2022}, here we adopt a simplified
metric to quantify the challenge. Let the observed width of a
cross-correlation function (CCF) be approximated by
\begin{equation}
  \sigma_{\rm obs} =
  \left(\sigma_{\rm inst}^2+\sigma_{\rm dyn}^2\right)^{1/2},
  \label{eq:sigmaobs}
\end{equation}
where $\sigma_{\rm inst}=c/(2.355R)$ is the Gaussian instrumental
width for resolving power $R$ (with $2.355 = 2\sqrt{2\ln 2}$ being the
ratio of the FWHM to the standard deviation of a Gaussian), and
$\sigma_{\rm dyn}$ is the intrinsic dynamical velocity-kernel width of
the atmospheric layer. For each apparent equatorial velocity $v_{\rm
  app}$, $\sigma_{\rm dyn}$ is computed as the flux-weighted standard
deviation of the disk-integrated velocity kernel at quadrature, which,
from numerical evaluation of Equation~(\ref{eq:kernel}) at quadrature,
is approximately $0.23\,v_{\rm app}$ for the Lambertian geometry
adopted here. The effective CCF signal-to-noise ratio required for a
3$\sigma$ detection of vertical shear between two layers is then
approximated as
\begin{equation}
  {\rm S/N}_{\rm eff} >
  \frac{3\sigma_{\rm obs,1}}{|\sigma_{\rm obs,1}-\sigma_{\rm obs,2}|},
  \label{eq:snr}
\end{equation}
where $\sigma_{\rm obs,1}$ refers to the upper (higher-velocity) layer.
This expression is not intended to replace an injection-recovery
analysis; rather, it illustrates the resolution dependence of the
problem. Because Equation~(\ref{eq:sigmaobs}) omits additional sources
of line broadening that will be present in practice, including the
intrinsic width of the incident stellar spectrum, template mismatch,
and pressure broadening of molecular features, the values shown in
Figure~\ref{fig:detectability} represent strict lower limits on the
required S/N. All of these omitted terms increase $\sigma_{\rm obs}$
and thereby dilute the fractional width difference between the two
atmospheric layers. The magnitude of the effect depends on how well
the stellar spectrum can be removed in the cross-correlation; for a
slowly rotating star with a well-matched template, the additional
broadening may be modest, but for rapidly rotating or spectrally
mismatched host stars the required S/N could increase by an order of
magnitude or more beyond the idealized values presented here.

Figure~\ref{fig:detectability} shows the required effective CCF
signal-to-noise ratio for detecting shear between a lower layer fixed
at 20~m~s$^{-1}$ and an upper layer with a range of apparent
velocities. The calculation shows why the terrestrial case is
difficult. At $R=10^5$, the instrumental width is much larger than the
100~m~s$^{-1}$ wind field, so the width difference is heavily diluted.
For a 100~m~s$^{-1}$ upper layer, the idealized effective CCF
signal-to-noise requirement is $\sim 2\times 10^4$ at $R=10^5$, $\sim
2\times 10^3$ at $R=3\times 10^5$, and $\sim 200$ at $R=10^6$. The
absolute values will change with line density and spectral model
assumptions, but the scaling indicates that sub-km~s$^{-1}$ shear
measurements favor very high spectral resolution and broad spectral
grasp. Proposed and under-development facilities operating at $R \sim
10^5$--$10^6$, including fiber-injection spectrographs on ELT-class
telescopes such as ELT/ANDES \citep{marconi2024a}, VLT/RISTRETTO
\citep{lovis2022}, and TMT/HROS \citep{froning2006}, and
post-coronagraphic high-dispersion coronagraphy \citep{wang2017a},
represent the instrumental frontier for such measurements.

The terrestrial case considered here is the most demanding application
of this diagnostic. The same physical effect operates in the reflected
light of gas giants, where it is considerably more accessible. The
reflected-light signal scales with the planet-to-star flux ratio,
which for a hot Jupiter at $\sim 0.05$~au with radius $\sim
10\,R_{\oplus}$ is of order $10^{-4}$, roughly four orders of
magnitude larger than for an exo-Venus in the HZ. Hot Jupiters and hot
Neptunes also exhibit much faster equatorial jets, with modeled and
measured wind speeds of several km~s$^{-1}$
\citep{showman2013a,brogi2016,seidel2020a}, so the dynamical velocity
width $\sigma_{\rm dyn}$ is one to two orders of magnitude larger than
the $\sim 20$--100~m~s$^{-1}$ terrestrial case. Both effects relax the
resolving-power and S/N requirements dramatically: for a
km~s$^{-1}$-scale wind field, the width difference between layers is
resolved at $R \sim 10^5$ rather than requiring $R \gtrsim 10^6$, and
the required effective CCF S/N drops by a similar factor. The recent
high-resolution detection of reflected light from the ultra-hot
exo-Neptune LTT-9779~b with ESPRESSO \citep{borsa2026}, which reported
a chromatic and longitudinal asymmetry in the reflected signal,
demonstrates that reflected-light cross-correlation spectroscopy of
close-in gas giants is now feasible. Such planets may therefore
provide the first observational tests of the wavelength-dependent
apparent-velocity diagnostic described here, well before it becomes
accessible for terrestrial worlds.

\section{Discussion}
\label{sec:disc}


\subsection{What Does a Reflected-Light Rotation Measurement Mean?}

The main result of this work is that reflected-light rotational
broadening should be interpreted as the velocity of the observable
layer, not automatically as the rotation of the solid planet. This
distinction is likely negligible for some rocky planets with optically
thin atmospheres and observable surfaces. It is central for Venus-like
planets, whose optically thick clouds obscure the lower atmosphere and
surface at visible wavelengths \citep{taylor2018,kane2021a}. In such
cases, the measurement is best described as a cloud-top velocity
measurement.

This interpretation is analogous to the lesson from Venus photometry
\citep{lee2020b}. A periodic brightness modulation can be real and
repeatable while still not being the solid-body rotation period. The
spectroscopic analog is that a broadening measurement can be real and
highly significant while still representing atmospheric superrotation.
The issue is a physical ambiguity in what layer is being observed
rather than an observational artifact. Analogous ambiguities have been
encountered in the interpretation of high-resolution spectroscopy of
transiting hot Jupiters, where vertical wind shear between the
day-night terminator and the photosphere produces both
wavelength-dependent velocity offsets and pressure-dependent line
broadening \citep{showman2013a,brogi2016,seidel2020a}.


\subsection{Combined Photometric and Spectroscopic Diagnostics}
\label{sec:combined}

A particularly powerful observational strategy is to combine
photometric period measurements with high-dispersion spectroscopic
velocity widths across multiple spectral regions. The Venus Akatsuki
results of \citet{lee2020b} demonstrate that disk-integrated
photometric periods of 3.7 and 4.6 days are detectable from the
superrotating atmosphere at UV and near-IR wavelengths. For an
exo-Venus analog, if photometric monitoring recovers a period of
$\sim 4$--5 days and high-resolution reflected-light spectra yield a
velocity width consistent with that same period, the planet could still
be either a rapid solid rotator or a Venus-like world. However, the two
scenarios make substantially different predictions about the wavelength and
pressure-level dependence of the measured velocity. A rapidly rotating
solid planet should produce an approximately constant apparent velocity
across all wavelengths and pressure levels probed, a photometric period
that is stable in time and wavelength-independent, and no systematic
change in the CCF width between high-opacity and low-opacity spectral
regions. A superrotating Venus-like atmosphere, by contrast, should
produce a systematic decrease in apparent velocity with increasing
sensitivity to deeper layers, approaching the true solid-body period of
hundreds of days for sub-cloud tracers, along with multiple photometric
periods associated with different dynamical modes at different altitudes
\citep{lee2020b} and wavelength-dependent CCF widths that track the
contribution function altitude of each spectral tracer.

Detecting even a factor of two difference in apparent velocity between
a near-UV cloud-deck tracer and a near-IR sub-cloud window would
constitute strong evidence for atmospheric superrotation. Such a
measurement, combined with a slow photometric variability period from
deeper layers, would be a definitive ``false spin'' detection and a
powerful comparative planetology result connecting to the atmospheric
angular momentum problem on Venus
\citep{read2018b,imamura2020,horinouchi2020}.


\subsection{How to Break the Degeneracy}

The most robust strategy is to measure apparent rotation as a function
of atmospheric depth. Different wavelength regions probe different
scattering altitudes in the atmosphere: for a Venus-like planet,
ultraviolet and visible reflected light are dominated by cloud and
haze properties at $\sim 60$--70~km altitude, whereas the
near-infrared thermal windows at 1.18, 1.74, and 2.3~$\mu$m provide
access to the sub-cloud atmosphere at 15--35~km altitude
\citep{meadows1996,taylor2018}.  A disagreement in apparent velocity
between a visible cloud-deck tracer and a near-IR sub-cloud window
would be a direct detection of vertical wind shear. Within a single
spectral bandpass, molecular line cores form higher in the atmosphere
than line wings, so comparing the CCF velocity from a strong-line core
to a weak line or line wing probes a range of altitude without
requiring multi-wavelength coverage.  Additionally, the reflected
stellar spectrum and the planetary molecular absorption spectrum need
not trace identical pressures, and a disagreement between their
inferred velocity fields would be a strong indicator of atmospheric
motion rather than solid-body rotation.

A complementary diagnostic is the comparison between photometric and
spectroscopic periods, as discussed in Section~\ref{sec:combined}. If
cloud patterns produce a photometric period of several days and
high-resolution reflected-light spectra produce a similar velocity
width, the planet may still be either rapidly rotating or Venus-like.
However, if deeper thermal or molecular features imply a much longer
period, the case for atmospheric superrotation becomes strong.
Conversely, if all tracers indicate the same velocity and the light
curve is stable across wavelengths, rapid solid-body rotation becomes
more plausible.

Phase-resolved line-profile morphology provides a third, if less
decisive, diagnostic. A solid-like wind profile can remain degenerate
with solid rotation at all phases, but real atmospheres need not obey
the $\cos\phi$ latitude dependence of solid-body rotation. Venus's
actual wind field departs significantly from $\cos\phi$ at
mid-latitudes and in the polar vortex regions
\citep{khatuntsev2013,horinouchi2020}, and such departures alter
line-profile asymmetries, centroid shifts, and phase-dependent widths
in ways that could eventually allow retrievals of the latitudinal wind
structure. These effects are illustrated by the non-solid-like wind case
in Figures~\ref{fig:phase_widths} and~\ref{fig:phase_centroids},
though realizing their diagnostic potential requires substantially
higher signal-to-noise ratio than a simple detection of broadening.


\subsection{Implications for HWO and Exo-Venus Science}

The results are directly relevant to precursor science for HWO. The
primary HWO goal of characterizing potentially habitable planets
\citep{stark2024b,tuchow2024,tuchow2025a} will
necessarily involve distinguishing Earth-like, Venus-like, and
intermediate terrestrial atmospheres \citep{kane2026a}. Venus analogs
are likely to be high-albedo targets in the visible because of their
cloud decks, but those same clouds complicate interpretation by
isolating the reflected spectrum from the surface. A rotation
measurement for such a world may therefore be a measurement of
atmospheric dynamics rather than of the solid planet.

Among the targets compiled by \citet{kane2026a} that lie within the
Venus Zone of their host stars \citep{kane2014e,ostberg2023a}, the
most promising candidates for even a photometric false-spin search are
those combining a bright host star, a favorable angular separation,
and a Venus-like insolation. The brightest such targets are the
super-Earths of nearby bright systems, including HD~20794~d ($V=4.3$,
separation $\sim 60$~mas), the HD~219134 planets ($V=5.6$, separations
$\sim 20$--36~mas), and GJ~411~b ($V=7.5$, at only 2.5~pc). For the
largest angular separations, and hence the most accessible to a
high-contrast architecture, the nearby M-dwarf systems are favorable,
including the Proxima Centauri and Barnard's Star planets, though
their fainter hosts ($V \gtrsim 9.5$) make high-dispersion velocity
work substantially more challenging. The temperate rocky planets in
these systems, with insolations of one to a few times the Earth value,
are plausible Venus analogs for which cloud formation and
superrotation are expected, and are therefore the natural first
targets for testing whether an apparent rotation signal originates in
the atmosphere rather than the surface. A full target-by-target yield
calculation, which would require an instrument simulator and
assumptions about each planet's atmospheric state, is beyond the scope
of this work but represents a valuable direction for HWO precursor
studies.

In practice, the observing architecture for detecting a false spin
would likely involve multiple facilities. Low- and moderate-resolution
HWO spectra and photometric monitoring could identify candidate
Venus-like planets and measure apparent rotation signals at the cloud
deck \citep{vaughan2023,stark2024a,morgan2024b}, and
moderate-dispersion spectroscopy from space has been shown to be a
promising route to characterizing directly imaged terrestrial planets
\citep{ruffio2026b}, while the vertical-shear diagnostic described
here would require high-dispersion coronagraphy from ELT-class
ground-based facilities or a future space-based high-resolution
spectroscopic mode. A practical advantage of the vertical-shear
diagnostic is that it does not require a temporal baseline: because it
relies on a wavelength-dependent velocity width rather than a temporal
modulation, it can in principle be performed from a single high-S/N
observation at a known orbital phase, as has been demonstrated for
directly imaged giant planets \citep{snellen2014b}. Photometric
rotation detection, by contrast, requires observations spanning
multiple apparent rotation periods to identify the periodic signal
\citep{palle2008,li2022a}.

This does not make the measurement less valuable. On the contrary,
detecting a false spin would provide evidence for a thick, dynamically
active atmosphere. A Venus-like apparent period of several days,
combined with a much slower lower-atmosphere or surface constraint,
would be a powerful comparative planetology result. It would connect
directly to the atmospheric angular momentum problem on Venus and to
the broader theory of superrotation in slowly rotating planetary
atmospheres \citep{read2018b,imamura2020}. The connection between
rotation rate and atmospheric properties extends beyond dynamics:
haze optical depth and cloud structure also vary with rotation rate
\citep{cohen2024b}, further complicating the interpretation of a
single broadening measurement. The challenge is to
avoid labeling the first measured velocity as the planet's rotation
period without testing for altitude dependence.


\subsection{Limitations and Future Work}

The model presented here is deliberately simple. The scattering is
Lambertian, the atmosphere is horizontally uniform, the spin axis is
not tilted, clouds are static, and the spectral contribution functions
are represented by discrete pressure levels. Real Venus-like planets
will have wavelength-dependent phase functions, patchy and evolving
clouds, nonzero obliquities, vertical and latitudinal wind shear,
thermal tides, planetary-scale waves, and potentially strong temporal
variability. These effects will complicate retrievals, but they do not
remove the fundamental degeneracy demonstrated by
Equation~(\ref{eq:degenerate}). If anything, the departures of real
Venus-like wind fields from the idealized $\cos\phi$ profile will
generally make the degeneracy easier to break, as discussed in
Section~\ref{sec:singlelayer}. The next step is to couple the velocity
model to realistic reflected and thermal spectra of Venus-like
exoplanets \citep{meadows1996,taylor2018,kane2026a}, including
wavelength-dependent cloud opacity, pressure-dependent molecular
contribution functions, and an instrument simulator for
high-dispersion coronagraphy
\citep{lovis2017,wang2017a}. Injection-recovery tests constructed from
synthetic observations of a Venus twin at several orbital phases would
determine the observing requirements for specific architectures,
including ELT-class ground-based facilities and future space-based
direct-imaging missions.


\section{Conclusions}
\label{sec:con}

The characterization of terrestrial exoplanet atmospheres through
direct imaging spectroscopy requires a careful accounting of what
layer is actually being observed. The results presented here
demonstrate that, for slowly rotating planets with Venus-like
superrotating atmospheres, reflected-light high-dispersion
spectroscopy will measure the velocity of the cloud deck rather than
that of the solid surface. For a single spectral tracer probing a
narrow range of pressures, a superrotating atmosphere whose zonal wind
follows a $\cos\phi$ latitude dependence is formally indistinguishable
from a rapidly rotating solid planet, producing an identical
disk-integrated velocity kernel. Because real Venus-like wind fields
depart significantly from this idealized form (exhibiting equatorial
jets, mid-latitude jets, and strong polar vortices) the exact
degeneracy represents a worst case, and the true diagnostic power of
high-resolution spectroscopy is correspondingly greater.

Venus provides the natural Solar System calibration point for this
effect. A Venus-like planet with a 243-day solid-body rotation period
and $\sim 100$~m~s$^{-1}$ cloud-top winds produces an apparent
rotation period of $\sim 4$--5 days in reflected light, consistent
with the 3.7 and 4.6 day photometric periods detected in
disk-integrated Akatsuki observations \citep{lee2020b}. The most
powerful diagnostic for breaking this degeneracy is vertical shear:
the inferred apparent rotation period should vary systematically with
wavelength, line strength, or molecular tracer if atmospheric
superrotation is responsible for the signal. Near-infrared thermal
windows at 1.18, 1.74, and 2.3~$\mu$m \citep{meadows1996,taylor2018}
offer access to sub-cloud layers where the contrast with cloud-top
velocities is greatest, though exploiting these emission features
would require extending the reflected-light formalism developed here
to a thermal-emission disk-weighting geometry. Phase-dependent
velocity centroids of order tens of m~s$^{-1}$ provide additional
information about the latitudinal wind structure and must be modeled
when interpreting reflected-light spectroscopy for orbital radial
velocity purposes. An idealized CCF-width analysis shows that
resolving 20--100~m~s$^{-1}$ vertical shear requires very high
spectral resolving power and high cross-correlation signal-to-noise
ratio, with ELT-class facilities operating at $R \sim 10^5$--$10^6$
\citep{snellen2015,lovis2017,wang2017a} representing the current
instrumental frontier for such measurements.

These results have direct implications for the preparation and
interpretation of HWO observations of Venus-like worlds. While HWO
itself is unlikely to achieve the spectral resolution required to
measure sub-km~s$^{-1}$ atmospheric velocities for terrestrial
planets, it will play the essential role of identifying the targets
for which the false-spin ambiguity arises by measuring apparent
rotation from photometric variability, constraining cloud-top albedos,
and flagging Venus-like candidates whose reflected spectra decouple
from the surface. The velocity measurements that resolve the ambiguity
would then be pursued with complementary high-dispersion
facilities. Future direct-imaging observations of terrestrial
exoplanets should report not only a single apparent rotation rate, but
the wavelength and pressure sensitivity of that measurement, along
with any dependence on orbital phase. For Venus-like worlds, the most
interesting discovery may be not the rotation of the planet, but the
false spin of its atmosphere.


\section*{Acknowledgements}

The author would like to thank the anonymous referee, whose feedback
helped to improve the manuscript. The author acknowledges useful
discussions within the Venus and exoplanet communities that motivated
the development of this problem. The results reported herein benefited
from collaborations and/or information exchange within NASA's Nexus
for Exoplanet System Science (NExSS) research coordination network
sponsored by NASA's Science Mission Directorate.




\end{document}